\documentclass[twocolumn,showpacs,preprintnumbers,amsmath]{revtex4}
\usepackage{amssymb}

\usepackage{graphicx}
\usepackage{amsmath}
\usepackage[rightcaption]{sidecap}

\begin{document}

\title{Non-equilibrium theory of charge qubit decoherence in the quantum point
contact measurement }
\author{\textbf{Ming-Tsung Lee$^{a}$ and Wei-Min Zhang$^{a,b}$}}
\date{\today}
\affiliation{$^{a}$National Center for Theoretical Science,
Tainan, Taiwan 70101, R.O.C. \\
$^{b}$Department of Physics and Center for Quantum Information
Science, National Cheng Kung University, Tainan, Taiwan 70101,
R.O.C.}

\begin{abstract}
A non-equilibrium theory describing the charge qubit dynamics
measured by a quantum point contact is developed based on
Schwinger-Keldysh's approach. Using the real-time diagram technique,
we derive the master equation to all orders in perturbation
expansions. The non-Markovian processes in the qubit dynamics is
naturally taken into account. The qubit decoherence, in particular,
the influence of the tunneling-electron fluctuation in the quantum
point contact with a longer time correlation, is studied in the
framework. We consider the Lorentzian-type spectral density to
characterize the channel mixture of the electron tunneling processes
induced by the measurement and determine the correlation time scale
of the tunneling-electron fluctuation. The result shows that as the
quantum point contact is casted with a narrower profile of the
spectral density, tunneling electrons can propagate with a longer
time correlation and lead to the non-Markovian processes of the
qubit dynamics. The qubit electron in the charge qubit will be
driven coherently. The quantum point contact measurement with the
minimum deviation of the electron tunneling processes prevents the
qubit state from the decoherence.
\end{abstract}

\pacs{03.65.Yz,73.23.-b,05.70.Ln,03.67.Lx} \maketitle

\section{INTRODUCTION}

With the rapid progress in nano-technology, the investigation of
quantum processes in systems coupling to a mesoscopical measurement
device becomes a very active research field in recent
years.\cite{nano1,nano2,nano3,nano4} Not only its practical
application to quantum communication and quantum
computation,\cite{qcomput} but also the theoretical interests in the
measurement-induced quantum decoherence \cite{nano3,tdecoh} have
attracted much attention. In particular, in the investigation of the
solid-state quantum computer with charge qubits, the quantum point
contact (QPC) has been served as an ultrasensitive electrometer
\cite{nano1,gurvitz,qpc1,qpc2,korotkov}. In the literature, the
measurement of charge qubit through the QPC has been treated based
on the so-called Markovian approximation,\cite{goan,stace,li,lee} in
which the time scale of the qubit dynamics is assumed much larger
than that of the tunneling-electron correlation in the QPC. This
theoretical treatment based on the Markovian approximation describes
only the qubit dynamics in the time-asymptotic quasi-equilibrium
state. Mesoscopically, the qubit decoherence occurs in the time
scale of the same order of the tunneling-electron correlation time
in the QPC, where the non-Markovian dynamics of the qubit is
significant. Thus, a non-equilibrium description to the qubit
dynamics is more desired. Recently, the considerations of the
solid-state system processing in the non-Markovian regime have
indeed received more attention. For instance, a local electron spin
coupling with a nuclear spin bath through the Fermi contact
hyperfine interaction in which the electron spin dynamics is in the
time scale shorter than the nuclear dipole-dipole correlation
time\cite{loss}, and electron transports of
interacting electron systems most likely involve the non-Markovian dynamics%
\cite{braggio,welack}. In our previous work\cite{lee}, the non-equilibrium
effect of the QPC on the qubit dynamics was studied by treating the fermi
energy fluctuation of the QPC reservoirs perturbatively. In this paper, a
fully non-equilibrium theory describing the qubit decoherence by the QPC
measurement is developed using Schwinger-Keldysh's approach\cite{schwinger}.

Historically, the Schwinger-Keldysh's approach is well developed to
systematically treat the non-equilibrium dynamics of a many-body
system.\cite {rammer,chou,schoeller1} For electron transports in
nano-devices, this approach has been used to study the transport
current fluctuation and the full counting statistics in the
single-electron transistor\cite {schoeller1,fcs}, the current
fluctuation in Kondo system\cite {wolfle,schoeller2,konig}, and the
noise spectrum in the QPC measurement of the charge
qubit\cite{shnirman}. To treat the non-equilibrium effect of the
electrical reservoir, literaturely the real-time diagrammatic
technique was constructed to diagrammatically calculate correlation
functions of the electrical reservoir order by
order.\cite{schoeller1,schoeller2} For the investigation of the
non-Markovian dynamics in our system, an alternative real-time
diagrammatic technique is developed. The master equation for the
charge qubit dynamics is derived and expressed in terms of all
orders irreducible diagrams to all orders. The non-Markovian
processes in the qubit dynamics can be fully taken into account. The
effect of the time variated reservoir fluctuation on the qubit
dynamics can then be explicitly studied in this formulism.

In addition, the assumption of the constant hopping amplitude of
tunneling electrons across the QPC barrier together with a constant
density of state of the QPC reservoirs is usually utilized to
specify the QPC structure related to the two reservoirs band
structure coupling to the two metal gates. It eventually leads to
the qubit dynamics in the Markovian limit.\cite{goan,stace,li,lee}
However, the QPC structure determines the correlation time scale of
the tunneling electron fluctuation in the QPC. The non-Markovian
processes of the qubit dynamics can emerge if a particular design of
the QPC structure is taken into account. We will consider in this
work a Lorentzian-type spectral density to characterize the
energy-level dependence of the hopping amplitude and the density of
state. A close relation between the qubit decoherence and the time
correlation of the tunneling-electron fluctuation shows that the
qubit decoherence can be controlled through the measurement
operation itself.

We organize the paper as follows: The theory of the charge qubit measurement
is presented in Sec. II. The real-time diagrammatic technique based on
Schwinger-Keldysh's approach are developed in Sec. III, where we also derive
the master equation for the reduced density operator of the charge qubit. In
Sec IV, the qubit dynamics is studied based on the master equation. The
influence of random electron-tunneling processes on the qubit decoherence is
illustrated in this section. Finally, a summary is given in Sec V.

\section{CHARGE QUBIT MEASUREMENT}

The charge qubit measurement using QPC is studied in the tunnel junction
regime\cite{gurvitz,korotkov,goan,shnirman,stace,li,lee}. In this regime,
the transmissions of all tunneling channels cross the QPC barrier are small
enough such that the electron tunneling becomes sensitive to the qubit
state. The qubit information can then be extracted from the output signal of
the QPC, while the backreaction of the measurement to the qubit states is
expected to be minimum. The Hamiltonian of the whole system is given by\cite
{gurvitz,korotkov,goan,shnirman,stace,li,lee}
\begin{equation}
H=H_{S}+H_{B}+H^{\prime },  \label{eq:total}
\end{equation}
where $H_{S}$ denotes the Hamiltonian of the charge qubit, $H_{B}$ the
Hamiltonian of the QPC with the electrical reservoirs consisting of the
source indexed by $l$ and the drain by $r$. The charge qubit state is
measured through the electron tunneling across the source and the drain. $%
H^{\prime }$ is the interaction Hamiltonian describing the electron
tunneling processes through the QPC with a qubit-state dependent hopping
amplitude $q_{rl}$. Explicitly,
\begin{equation}
H_{S}=\frac{\epsilon }{2}\sigma _{z}+\frac{\vartriangle }{2}\sigma _{x},
\label{eq:sysh}
\end{equation}
\begin{equation}
H_{B}=\sum_{l}\epsilon _{l}a_{l}^{+}a_{l}+\sum_{r}\epsilon
_{r}a_{r}^{+}a_{r},
\end{equation}
\begin{equation}
H^{\prime }=\sum_{rl}(q_{rl}a_{r}^{+}a_{l}+q_{lr}a_{l}^{+}a_{r}).
\end{equation}

We shall formulate the non-equilibrium theory of the electron tunnelings
coupled with the qubit dynamics in the electron-hole representation, in
which $H_{B}$ and $H^{\prime }$ can be written equivalently as

\begin{eqnarray}
H_{B} &=&\sum_{l>0}(\epsilon _{l}^{e}\alpha _{l}^{+}\alpha _{l}+\epsilon
_{l}^{h}\beta _{l}^{+}\beta _{l})  \nonumber \\
&&+\sum_{r>0}(\epsilon _{r}^{e}\alpha _{r}^{+}\alpha _{r}+\epsilon
_{r}^{h}\beta _{r}^{+}\beta _{r}),  \label{eq:bath}
\end{eqnarray}
\begin{eqnarray}
H^{\prime } &=&\sum_{rl>0}q_{rl}(\alpha _{r}^{+}\alpha _{l}+\beta _{r}\beta
_{l}^{+}+\beta _{r}\alpha _{l}  \nonumber \\
&&+\alpha _{r}^{+}\beta _{l}^{+})+\text{H.c.,}  \label{eq:interaction}
\end{eqnarray}
where $\alpha _{l,r}^{+}$($\alpha _{l,r}$) and $\beta _{l,r}^{+}$($\beta
_{l,r}$) are respectively the creation (annihilation) operators of the
electron and hole, the corresponding electron and hole energies $\epsilon
_{l,r}^{e}=\epsilon _{l,r}-\mu _{L,R}$ and $\epsilon _{l,r}^{h}=\mu
_{L,R}-\epsilon _{l,r}$ with respect to the chemical potential $\mu _{L}$ of
the source and $\mu _{R}$ of the drain. Since the coupling $q_{rl}$ depends
on the qubit state, it is indeed a coupling function of the qubit operator.
Meanwhile, $q_{rl}$ also depends on the measurement device structure. We
remain the discussion of the QPC measurement to the qubit decoherence with a
practical $q_{rl}$ later.

The qubit dynamics is determined by the master equation for the reduced
density operator $\rho _{S}(t)=$Tr$_{B}\left[ \rho _{tot}(t)\right] $, where
$\rho _{tot}(t)$ is the total density operator of the whole system, and the
partial trace Tr$_{B}$ integrates over all the degrees of freedom of the QPC
reservoir. From the Liouville equation $\frac{\partial }{\partial t}\rho
_{tot}=-i\left[ H,\rho _{tot}\right] $ for the total density operator, it
can be shown that the reduced density operator $\rho _{S}(t)$ obeys the
following equation of motion (the master equation) \cite
{eom,welack,spectrum1}
\begin{eqnarray}
&&\frac{\partial }{\partial t}\rho _{S}(t)=-i\left[ H_{S},\rho
_{S}(t)\right] -i\text{Tr}_{B}\Big[L^{\prime }Q_{H_{0}}(t-t_{0})  \nonumber
\\
&&~~~~~~~~~~\times \rho _{tot}(t_{0})\Big]-\int_{t_{0}}^{t}d\tau K(t-\tau
)*\rho _{S}(\tau ),  \label{eq:eoms}
\end{eqnarray}
where $K(t-\tau )*\rho _{S}(\tau )\equiv $Tr$_{B}[L^{\prime
}Q_{H_{0}}(t-\tau )L^{\prime }\rho _{tot}(\tau )]$, $L^{\prime }\equiv
\left[ H^{\prime },\right] $, $Q_{H_{0}}(t)\ =e^{-it\left[
H_{S}+H_{B},\right] }$, and $t_{0}$ is the initial time that the interaction
between the qubit and the QPC measurement turns on. The derivation of Eq. (%
\ref{eq:eoms}) can be found in Appendix A. The second term in the above
equation has no contribution due to the particle number conservation in
electron tunneling processes. Since the QPC output signal records the
information of the qubit state through the interaction between the QPC and
the qubit, the induced backreaction from the fluctuant reservoirs will
result in qubit decoherence which is in principle non-Markovian. The
non-Markovian dynamics of the qubit is described by the term $%
-\int_{t_{0}}^{t}d\tau K(t-\tau )*\rho _{S}(\tau )$, which contains all
influences of the fluctuant reservoir to the qubit dynamics. The qubit
decoherence is thus totally governed by Eq. (\ref{eq:eoms}). In the
following discussion, we will derive the master equation for the reduced
density operator in terms of a diagrammatic perturbation expansion of Eq. (%
\ref{eq:eoms}).

\section{REAL-TIME DIAGRAMMATIC TECHNIQUE}

We begin with the Schwinger-Keldysh's approach\cite{schwinger} to explicitly
calculate the term $-\int_{t_{0}}^{t}d\tau K(t-\tau )*\rho _{S}(\tau )$ in
Eq. (\ref{eq:eoms}). In the interaction picture, the interaction Hamiltonian
Eq. (\ref{eq:interaction}) can be rewritten as
\begin{equation}
\hat{H}_{t}^{\prime }=\sum_{rl>0}\Big\{\hat{q}_{rl}(t)\hat{\psi}_{r}^{+}(t)%
\hat{\psi}_{l}(t)+\hat{q}_{rl}^{+}(t)\hat{\psi}_{l}^{+}(t)\hat{\psi}_{r}(t)%
\Big\},  \label{eq:intaction1}
\end{equation}
where the field operator $\hat{\psi}_{k}(t)$ is defined by
\begin{equation}
\hat{\psi}_{k}(t)=\alpha _{k}e^{-i\epsilon _{k}^{e}(t-t_{0})}+\beta
_{k}^{+}e^{i\epsilon _{k}^{h}(t-t_{0})}
\end{equation}
with the index $k=l,r$ labeled respectively for the source and the drain. In
addition, $K(t-\tau )*\rho _{S}(\tau )$ in the interaction picture can be
rewritten as
\begin{eqnarray}
&&K(t-\tau )*\rho _{S}(\tau )=Q_{S}(t-t_{0})  \nonumber \\
&&\times \Big(\text{Tr}_{B}\left[ \hat{H}_{t}^{\prime }\hat{H}_{\tau
}^{\prime }T_{\tilde{p}}\left\{ e^{-i\int_{\tilde{p},t_{0}}^{\tau }ds\hat{H}%
_{s}^{\prime }}\rho _{tot}(t_{0})\right\} \right]   \nonumber \\
&&-\text{Tr}_{B}\left[ \hat{H}_{t}^{\prime }T_{\tilde{p}}\left\{ e^{-i\int_{%
\tilde{p},t_{0}}^{\tau }ds\hat{H}_{s}^{\prime }}\rho _{tot}(t_{0})\right\}
\hat{H}_{\tau }^{\prime }\right]   \nonumber \\
&&-\text{Tr}_{B}\left[ \hat{H}_{\tau }^{\prime }T_{\tilde{p}}\left\{
e^{-i\int_{\tilde{p},t_{0}}^{\tau }ds\hat{H}_{s}^{\prime }}\rho
_{tot}(t_{0})\right\} \hat{H}_{t}^{\prime }\right]   \nonumber \\
&&+\text{Tr}_{B}\left[ T_{\tilde{p}}\left\{ e^{-i\int_{\tilde{p}%
,t_{0}}^{\tau }ds\hat{H}_{s}^{\prime }}\rho _{tot}(t_{0})\right\} \hat{H}%
_{\tau }^{\prime }\hat{H}_{t}^{\prime }\right] \Big),  \label{eq:kernali}
\end{eqnarray}
where $Q_{S}(t)=e^{-it\left[ H_{S},\right] }$ and $\int_{\tilde{p}%
,t_{0}}^{t}ds$ denotes the closed-time-path integral along the
closed-time-path contour $\tilde{p}$ \cite{schwinger,rammer,chou,schoeller1}
with the range of the real time axis from $t_{0}$ to $t$. The closed time
path contains the positive branch and the negative branch. The positive
branch coincides with the real time axis, and the negative branch is
reversed with respect to the real time axis. Taking perturbation expansion
of Eq. (\ref{eq:kernali})
\begin{eqnarray}
&&K(t-\tau )*\rho _{S}(\tau )=K^{(0)}(t-\tau )*\rho _{S}(\tau )
\nonumber \\
&&~~~~~~~~~~~~~~~+K^{(1)}(t-\tau )*\rho _{S}(\tau)+\cdots
\label{eq:ke}
\end{eqnarray}
through the expansion\cite{rammer}
\begin{eqnarray}
&&T_{\tilde{p}}\left\{ e^{-i\int_{\tilde{p}}ds\hat{H}_{s}^{\prime
}}\rho _{tot}(t_{0})\right\}=\sum_{n=0}^{\infty }\frac{(-i)^{n}}{n!}\int_{\tilde{p}}ds_{1}\cdots \int_{%
\tilde{p}}ds_{n}  \nonumber \\
&&~~~~~~~~~~~~~~~~~~~~~~~\times T_{\tilde{p}}\left\{ \hat{H}_{s_{1}}^{\prime }\cdots \hat{H}%
_{s_{n}}^{\prime }\rho _{tot}(t_{0})\right\} , \label{eq:expan}
\end{eqnarray}
we can diagrammatically illustrate Eq. (\ref{eq:kernali}) as
\begin{figure}[tbph]
\includegraphics*[angle=270,scale=.375]{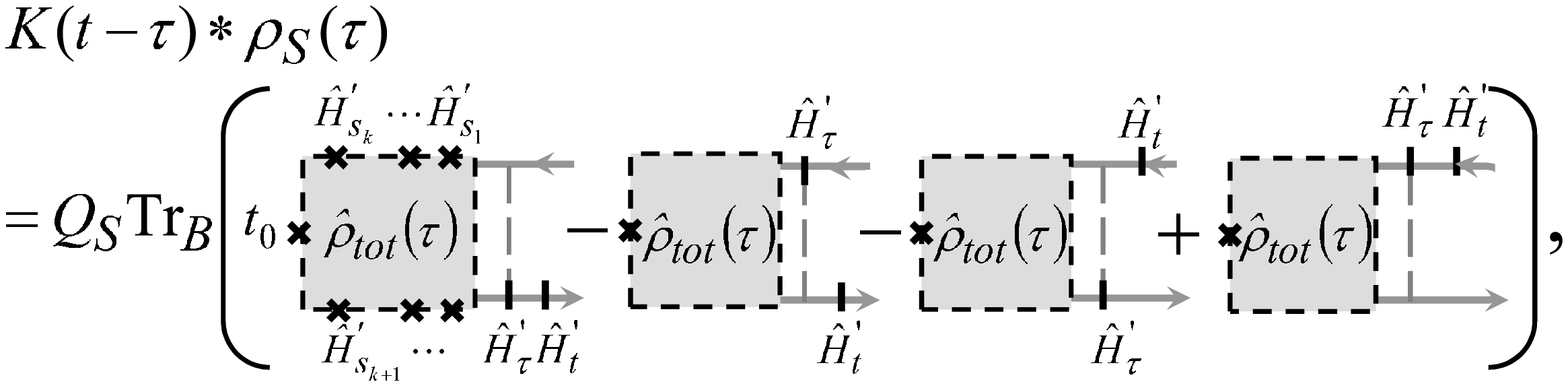}
\end{figure}
\begin{equation}
\textrm{with} \quad \hat{\rho}_{tot}\left( \tau
\right)=T_{\tilde{p}}\left\{ e^{-i\int_{\tilde{p},t_{0}}^{\tau
}ds\hat{H}_{s}^{\prime }}\rho_{tot}\left(t_{0}\right) \right\}. \\
\end{equation}

According to the Wick's theorem, all higher order correlation
functions of the fermion field operators can be built up in terms of
the unperturbed Green's function.\cite{rammer,chou,schoeller1} The
reservoir contour-ordered Green's function is defined by
\begin{equation}
\tilde{G}_{k}(t,t^{\prime })=-i\left\langle T_{\tilde{p}}\Big(\hat{\psi}%
_{k}(t)\hat{\psi}_{k}^{+}(t^{\prime })\Big)\right\rangle ,  \label{eq:tcgf}
\end{equation}
where the contour-ordering operator $T_{\tilde{p}}$ orders the operator $%
\hat{\psi}_{k}(t)\hat{\psi}_{k}^{+}(t^{\prime })$ according to their time
arguments along the $\tilde{p}$, and $\left\langle .\right\rangle $ denotes
the non-equilibrium statistical average. Explicitly, $\tilde{G}%
_{k}(t,t^{\prime })$ contains four components:
\begin{equation}
\tilde{G}_{k}(t,t^{\prime })=\left(
\begin{array}{ll}
G_{F,k}(t,t^{\prime }) & G_{<,k}(t,t^{\prime }) \\
G_{>,k}(t,t^{\prime }) & G_{\tilde{F},k}(t,t^{\prime })
\end{array}
\right) ,  \label{eq:greenm}
\end{equation}
i.e. the time-ordered Green's function $G_{F,k}$, the anti-time-ordered
Green's function $G_{\tilde{F},k}$, the correlation Green's functions $%
G_{<,k}(t,t^{\prime })$ and $G_{>,k}$, defined respectively as
\begin{equation}
G_{F,k}(t,t^{\prime })=-i\left\langle T(\hat{\psi}_{k}(t)\hat{\psi}%
_{k}^{+}(t^{\prime }))\right\rangle ,
\end{equation}
\begin{equation}
G_{\tilde{F},k}(t,t^{\prime })=-i\left\langle \tilde{T}(\hat{\psi}_{k}(t)%
\hat{\psi}_{k}^{+}(t^{\prime }))\right\rangle ,
\end{equation}
\begin{equation}
G_{<,k}(t,t^{\prime })=i\left\langle \hat{\psi}_{k}^{+}(t^{\prime })\hat{%
\psi}_{k}(t)\right\rangle ,
\end{equation}
\begin{equation}
G_{>,k}(t,t^{\prime })=-i\left\langle \hat{\psi}_{k}(t)\hat{\psi}%
_{k}^{+}(t^{\prime })\right\rangle ,
\end{equation}
where $T$ is the normal time-ordering operator and $\tilde{T}$ the
anti-time-ordering operator. The unperturbed reservoir
contour-ordered Green's functions are easy calculated,
\begin{eqnarray}
-iG_{<,k}^{(0)}(t,t^{\prime }) &=&\text{Tr}_{B}\Big[ \hat{\psi}%
_{k}^{+}(t^{\prime })\hat{\psi}_{k}(t)\rho _{B}^{(0)}\Big]  \nonumber \\
&=&\Big(1-f(\epsilon _{k}^{h})\Big)e^{i\epsilon _{k}^{h}(t-t^{\prime })}
\nonumber \\
&&+f(\epsilon _{k}^{e})e^{-i\epsilon _{k}^{e}(t-t^{\prime })},
\label{eq:0gfl}
\end{eqnarray}
\begin{eqnarray}
iG_{>,k}^{(0)}(t,t^{\prime }) &=&\text{Tr}_{B}\Big[ \hat{\psi}_{k}(t)\hat{%
\psi}_{k}^{+}(t^{\prime })\rho _{B}^{(0)}\Big]  \nonumber \\
&=&\Big(1-f(\epsilon _{k}^{e})\Big)e^{-i\epsilon _{k}^{e}(t-t^{\prime })}
\nonumber \\
&&+f(\epsilon _{k}^{h})e^{i\epsilon _{k}^{h}(t-t^{\prime })},
\label{eq:0gfg}
\end{eqnarray}
\begin{equation}
G_{F,k}^{(0)}(t,t^{\prime })=\theta (t-t^{\prime })G_{>,k}^{(0)}(t,t^{\prime
})+\theta (t^{\prime }-t)G_{<,k}^{(0)}(t,t^{\prime }),  \label{eq:0gft}
\end{equation}
\begin{equation}
G_{\tilde{F},k}^{(0)}(t,t^{\prime })=\theta (t-t^{\prime
})G_{<,k}^{(0)}(t,t^{\prime })+\theta (t^{\prime
}-t)G_{>,k}^{(0)}(t,t^{\prime }),  \label{eq:0gfat}
\end{equation}
Here, the electrical reservoirs are assumed to be in the thermal equilibrium
state $\rho _{B}^{(0)}$ initially (at $t_{0}$), and $f(\epsilon )$ is the
Fermi-Dirac distribution function. These Green's functions describe the
contractions of the unperturbed fermion pair with four different time
contour-orderings.

\subsection{The diagrammatic rules}

\begin{figure}[tbph]
\includegraphics*[angle=270,scale=.35]{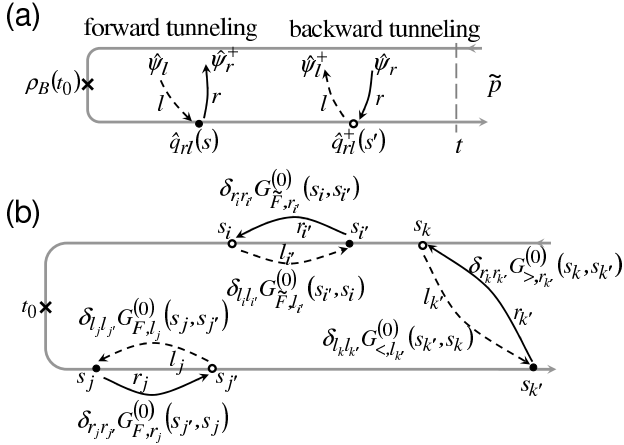}
\caption{The diagrammatic representation. (a) The forward tunneling vertex $%
\hat{q}_{rl}(t)\hat{\psi}_{r}^{+}(t)\hat{\psi}_{l}(t)$ and the backward
tunneling vertex $\hat{q}_{lr}(t)\hat{\psi}_{l}^{+}(t)\hat{\psi}_{r}(t)$.
(b) The free propagators.}
\label{resvrline}
\end{figure}

In order to systematically trace out the reservoir degrees of freedom to all
orders in the perturbation expansion with a correct operator ordering for
the remaining degrees of freedom of the qubit, we shall use the real-time
diagrammatic expansion with the diagrammatic rules defined as follows:
According to the interaction Hamiltonian of Eq. (\ref{eq:intaction1}), two
kinds of the tunnelings, the forward and the backward tunneling, across the
QPC barrier are involved. The diagrammatic representation of the forward
tunneling vertex $\hat{q}_{rl}\hat{\psi}_{r}^{+}\hat{\psi}_{l}$ and the
backward tunneling vertex $\hat{q}_{rl}^{+}\hat{\psi}_{l}^{+}\hat{\psi}_{r}$
are depicted in Fig. \ref{resvrline} (a). For the forward tunneling vertex,
the incoming dashed line labeled by $l$ represents the electrons (or holes)
above (below) the chemical potential $\mu _{L}$ are destroyed (created) in
the source, and the outgoing solid line labeled by $r$ represents the
electrons (or holes) with the energy level above (below) the chemical
potential $\mu _{R}$ are created (destroyed) in the drain. The QPC current
due to this tunneling effectively flows from the source to the drain. The
coupling operator $\hat{q}_{rl}$ is presented in the vertex with a filled
circle for the forward tunneling. Similarly, the interaction associated with
the backward tunneling is depicted by the incoming solid line labeled by $r$
and the outgoing dashed line labeled by $l$, and the coupling operator $\hat{%
q}_{rl}^{+}$ presented in the vertex is denoted with a hollow circle.

Meanwhile, all vertices in the expansion should be connected in pairs by the
electron propagators. There are only two types of the propagators involved:
the solid line, starting from the vertex $\hat{q}_{rl}\hat{\psi}_{r}^{+}\hat{%
\psi}_{l}$ to the vertex $\hat{q}_{rl}^{+}\hat{\psi}_{l}^{+}\hat{\psi}_{r}$,
represents the electron propagates in the drain, and the dash line, starting
from the vertex $\hat{q}_{rl}^{+}\hat{\psi}_{l}^{+}\hat{\psi}_{r}$ to the
vertex $\hat{q}_{rl}\hat{\psi}_{r}^{+}\hat{\psi}_{l}$, is the propagator in
the source. These propagators connecting two vertices with both the time
arguments $s$ and $s^{\prime }$ located at the positive (negative) branch of
the closed time path represent the (anti-) time-ordered Green's function $%
G_{F(\tilde{F})}^{(0)}(s,s^{\prime })$. The arrow of the propagator
coincides with the propagating direction of the tunneling electron. While
the propagator connecting two vertices, one located at the positive branch
and the other at the negative branch separately, represents the correlation
Green's function $G_{<(>)}^{(0)}(s,s^{\prime })$ with respect to the arrow
pointing to the vertex at the positive (negative) branch. The diagrammatic
representation of the free propagators is summarized in Fig. \ref{resvrline}
(b). Different from the usual closed-time-path contour used in the
literature, we choose an alternate contour $\tilde{p}$ depicted in Fig. \ref
{resvrline}. The lower (upper) axis represents the positive (negative) time
branch of the closed-time-path contour. The contour ordering $t\geq _{\tilde{%
p}}t^{\prime }$ denotes the time $t^{\prime }$ is former than $t$ along the
arrow of the contour $\tilde{p}$, and the corresponding operator $\hat{q}%
(t^{\prime })$ is applied earlier than $\hat{q}(t)$. This prescription makes
the calculation more convenient in the treatment of the time ordering of
coupling operators discussed below.

\subsection{The real-time diagrammatic expansion}

Now, we can diagrammatically calculate the partial trace in the expansion of
Eq. (\ref{eq:kernali})
\begin{equation}
\text{Tr}_{B}\left[ T_{\tilde{p}}\left\{ \hat{H}_{t}^{\prime }\hat{H}_{\tau
}^{\prime }\hat{H}_{s_{1}}^{\prime }\cdots \hat{H}_{s_{n}}^{\prime }\rho
_{tot}(t_{0})\right\} \right] ,  \label{eq:ptrace1}
\end{equation}
where $\hat{H}_{s_{1}}^{\prime }\cdots \hat{H}_{s_{n}}^{\prime }$ comes from
the perturbation expansion of the evolution operator $T_{\tilde{p}}\{\exp
(-i\int_{\tilde{p},t_{0}}^{\tau }ds\hat{H}_{s}^{\prime })\cdot \}$ with $%
t\geq \tau $ $\geq $ $(s_{1},\cdots ,s_{n})$. Conveniently, the vertex with
the time argument $s_{i=1,\cdots ,n}$ is called the internal vertex and that
with $(t,\tau )$ is called the external vertex. The time contour ordering $%
T_{\tilde{p}}$ in Eq. (\ref{eq:ptrace1}) comprises all the permutation of
the time series $\{t\geq \tau \geq (s_{1},\cdots ,s_{n})\}$ along the closed
time path $\tilde{p}$, namely, one must sum all allowed time contour
orderings $T_{\tilde{p}}=\sum_{i}T_{\tilde{p},i}$. In terms of $\hat{q}_{rl}%
\hat{\psi}_{r}^{+}\hat{\psi}_{l}$ and $\hat{q}_{rl}^{+}\hat{\psi}_{l}^{+}%
\hat{\psi}_{r}$, each component of Eq. (\ref{eq:ptrace1}) can be expressed
as
\begin{eqnarray}
&&\text{Tr}_{B}\Big[T_{\tilde{p}}\Big\{\Big(\hat{q}_{rl}(t)\hat{\psi}%
_{r}^{+}(t)\hat{\psi}_{l}(t)\Big)^{(+)}  \nonumber \\
&&~~~~~~\times \Big(\hat{q}_{r^{\prime }l^{\prime }}(\tau )\hat{\psi}%
_{r^{\prime }}^{+}(\tau )\hat{\psi}_{l^{\prime }}(\tau )\Big)^{(+)}\cdots
\nonumber \\
&&~~~~~~\times \Big(\hat{q}_{r_{n}l_{n}}(s_{n})\hat{\psi}_{r_{n}}^{+}(s_{n})%
\hat{\psi}_{l_{n}}(s_{n})\Big)^{(+)}\rho _{tot}(t_{0})\Big\}\Big]  \nonumber
\\
&=&\sum_{i}\mathcal{E}_{rlr^{\prime }l^{\prime }\cdots r_{n}l_{n};i}\mathcal{%
S}_{rlr^{\prime }l^{\prime }\cdots r_{n}l_{n};i},  \label{eq:pe}
\end{eqnarray}
where $\mathcal{S}_{rlr^{\prime }l^{\prime }\cdots r_{n}l_{n};i}$ associated
with coupling operators is defined as
\begin{eqnarray}
&&\mathcal{S}_{rlr^{\prime }l^{\prime }\cdots r_{n}l_{n};i}(t,\tau
,s_{1},\cdots ,s_{n})  \nonumber  \label{eq:ptrq} \\
&\equiv &T_{\tilde{p},_{i}}\left\{ \hat{q}_{rl}^{(+)}(t)\hat{q}_{r^{\prime
}l^{\prime }}^{(+)}(\tau )\hat{q}_{r_{1}l_{1}}^{(+)}(s_{1})\cdots \hat{q}%
_{r_{n}l_{n}}^{(+)}(s_{n})\rho _{S}(t_{0})\right\} ,  \nonumber \\
&&
\end{eqnarray}
and the corresponding coefficient $\mathcal{E}_{rlr^{\prime }l^{\prime
}\cdots r_{n}l_{n};i}$ is the contribution of integrating out the electron
reservoirs and is defined by
\begin{eqnarray}
&&\mathcal{E}_{rlr^{\prime }l^{\prime }\cdots r_{n}l_{n};i}(t,\tau
,s_{1},\cdots ,s_{n})  \nonumber \\
&\equiv &\text{Tr}_{B}\Big[T_{\tilde{p},i}\Big\{(\hat{\psi}_{r}^{+}(t)\hat{%
\psi}_{l}(t))^{(+)}(\hat{\psi}_{r^{\prime }}^{+}(\tau )\hat{\psi}_{l^{\prime
}}(\tau ))^{(+)}\cdots  \nonumber \\
&&~~~~~~~~~~~~~~\times (\hat{\psi}_{r_{n}}^{+}(s_{n})\hat{\psi}%
_{l_{n}}(s_{n}))^{(+)}\rho _{B}^{(0)}\Big\}\Big].  \label{eq:ptrcoef}
\end{eqnarray}
Note that $\mathcal{S}_{rl\cdots ;i}$ in Eq. (\ref{eq:ptrq}) consists of $%
n+2 $ vertices which is ordered according to $T_{\tilde{p},i}$. Due to the
particle number conservation in electron tunneling processes, only even
orders ($n=$even) in the perturbation expansion have contribution. Half of
these $n+2$ vortices will carry with the coupling operator $\hat{q}^{+}$,
and the others with $\hat{q}$. Each $\mathcal{E}_{rl\cdots ;i}\mathcal{S}%
_{rl\cdots ;i}$ is expressed by a set of topology-independent diagrams, in
which each diagram are composed of several allowed closed loops connecting $%
n+2$ vertices. The topology-independence means the order and the direction
of all propagators lines and the loop assembly are different. The
coefficient $\mathcal{E}_{rl\cdots ;i}$ can be directly calculated only from
this set of topology-independent diagrams by summing all the
topology-independent diagrams with a prefactor $(-1)^{(n+2)/2+l}$, where $l$
is the loop number in the individual topology-independent diagram. The
prefactor $(-1)^{(n+2)/2}$ comes from the factor $\left( -i\right) ^{(n+2)}$
of the $n$-th order perturbation, and the prefactor $(-1)^{l}$ is due to the
permutation between the fermion operators in the contraction. An explicit
example of calculating $\mathcal{E}_{rl\cdots ;i}\mathcal{S}_{rl\cdots ;i}$
for the order of $n=2$ can be found in Appendix B.

Accordingly, we introduce a loop operator $\mathcal{\hat{L}}_{\tau }(\cdots
) $ to calculate the total contribution of all time contour orderings in the
$n $-th order perturbation $K^{(n)}(t-\tau )*\rho _{S}(\tau )$. This loop
operator is defined by a loop in particular topology-independent diagrams
along \textit{the real time axis},
\begin{eqnarray}
&&\mathcal{\hat{L}}_{\tau }\left(
\begin{array}{lllllllllll}
s_{1} &  & s_{2} &  & s_{3} &  & \cdots s_{i}\cdots &  & s_{2p} &  & s_{1}
\\
& k_{1} &  & k_{1}^{\prime } &  & k_{2} & \cdots k_{j}\cdots & k_{p} &  &
k_{p}^{\prime } &
\end{array}
\right)  \nonumber \\
&\equiv &\sum\limits_{k_{1}\cdots k_{p}}\sum\limits_{k_{1}^{\prime }\cdots
k_{p}^{\prime }}\int_{t_{0}}^{\tau }ds_{1}\cdots \int_{t_{0}}^{\tau }ds_{2p}
\nonumber \\
&&\times \text{Tr}\Big[D[\alpha _{k_{1}}(s_{1})]\tilde{G}%
_{k_{1}}^{(0)}(s_{1},s_{2})D[\alpha _{k_{1}^{\prime }}(s_{2})]\tilde{G}%
_{k_{1}^{\prime }}^{(0)}(s_{2},s_{3})  \nonumber \\
&&\cdots D[\alpha _{k_{p}^{\prime }}(s_{2p})]\tilde{G}_{k_{p}^{\prime
}}^{(0)}(s_{2p},s_{1})\Big],  \label{eq:loopo}
\end{eqnarray}
where $s_{i=1,\cdots ,2p}\leq \tau $ are the time arguments of the internal
vertices, $p$ is an integer with $2p\leq n$, $k_{j}$ and $k_{j}^{\prime }$
are the energy indexes with $(k_{j}=l_{j},k_{j}^{\prime }=r_{j})$ or $%
(k_{j}=r_{j},k_{j}^{\prime }=l_{j})$, $\tilde{G}_{k}^{(0)}(s,s^{\prime })$
is the Keldysh's matrix shown in Eq. (\ref{eq:greenm}), and the $2\times 2$
functional-derivative matrix $D[\alpha (s)]$ is defined by $D[\alpha
(s)]=\left(
\begin{array}{ll}
\frac{\delta }{\delta \alpha (s)} & 0 \\
0 & \frac{\delta }{\delta \alpha ^{*}(s)}
\end{array}
\right) $ with $\alpha (s)$ being a time-dependent parameter. The loop
involving the external vertices is denoted as $\mathcal{\hat{L}}_{\tau
}\left(
\begin{array}{lllllll}
t &  & s_{1} & \cdots \tau \cdots & s_{2p-2} &  & t \\
& k_{1} &  & \cdots k_{j}\cdots &  & k_{p}^{\prime } &
\end{array}
\right) $. This definition is the same as Eq. (\ref{eq:loopo}) but without
taking the time integral for the time arguments $t$ and $\tau $. The loop
operator is written in such a way that a pair of vertices labeled by the
time arguments $s^{\prime }$ and $s$ together with the propagator with the
energy index $k$ represent a segment of the loop operator $\mathcal{\hat{L}}%
_{\tau }\left(
\begin{array}{lll}
\cdots s^{\prime } &  & s\cdots \\
& k &
\end{array}
\right) $. Each internal (external) vertex with the coupling operator $\hat{q%
}_{rl}$ or $\hat{q}_{rl}^{+}$ is described by the operator $D[\alpha _{l}]$ $%
(D[\gamma _{l}])$ and $D[\alpha _{r}]$ $(D[\gamma _{r}])$, respectively. The
connecting propagator is the Keldysh's matrix $\tilde{G}_{k}^{(0)}(s,s^{%
\prime })$. Along the propagating direction in each loop, the corresponding
loop operator is written down in the order from right to left. The loop is
end-point-independent, with which the definition of the loop operator
coincides. Since the hermitian of the physical quantity, $K^{(n)}(t-\tau
)*\rho _{S}(\tau )$ can be separated into two parts which are hermitian
conjugate to each other. This leads to two sets of the diagrams which are
dual each other. The duality of two diagrams is defined that the
replacements of the vertices $\hat{q}$ ($\hat{q}^{+}$) in one diagram by the
vertices $\hat{q}^{+}$ ($\hat{q}$) equals to each other. Therefore, we only
need to calculate one set of the diagrams. In terms of the loop operators,
only topology-independent diagrams along the real time axis should be taken
into account. Furthermore, besides the prefactor $\frac{1}{n!}%
(-1)^{(n+2)/2+l}$, where the factor $\frac{1}{n!}$ comes from the $n$-th
order perturbation, a weight factor (the number of topology-equivalent
diagrams) should also be added for each topology-independent diagram to
calculate $K^{(n)}(t-\tau )*\rho _{S}(\tau )$ correctly.

\begin{figure}[tbph]
\includegraphics*[angle=270,scale=.36]{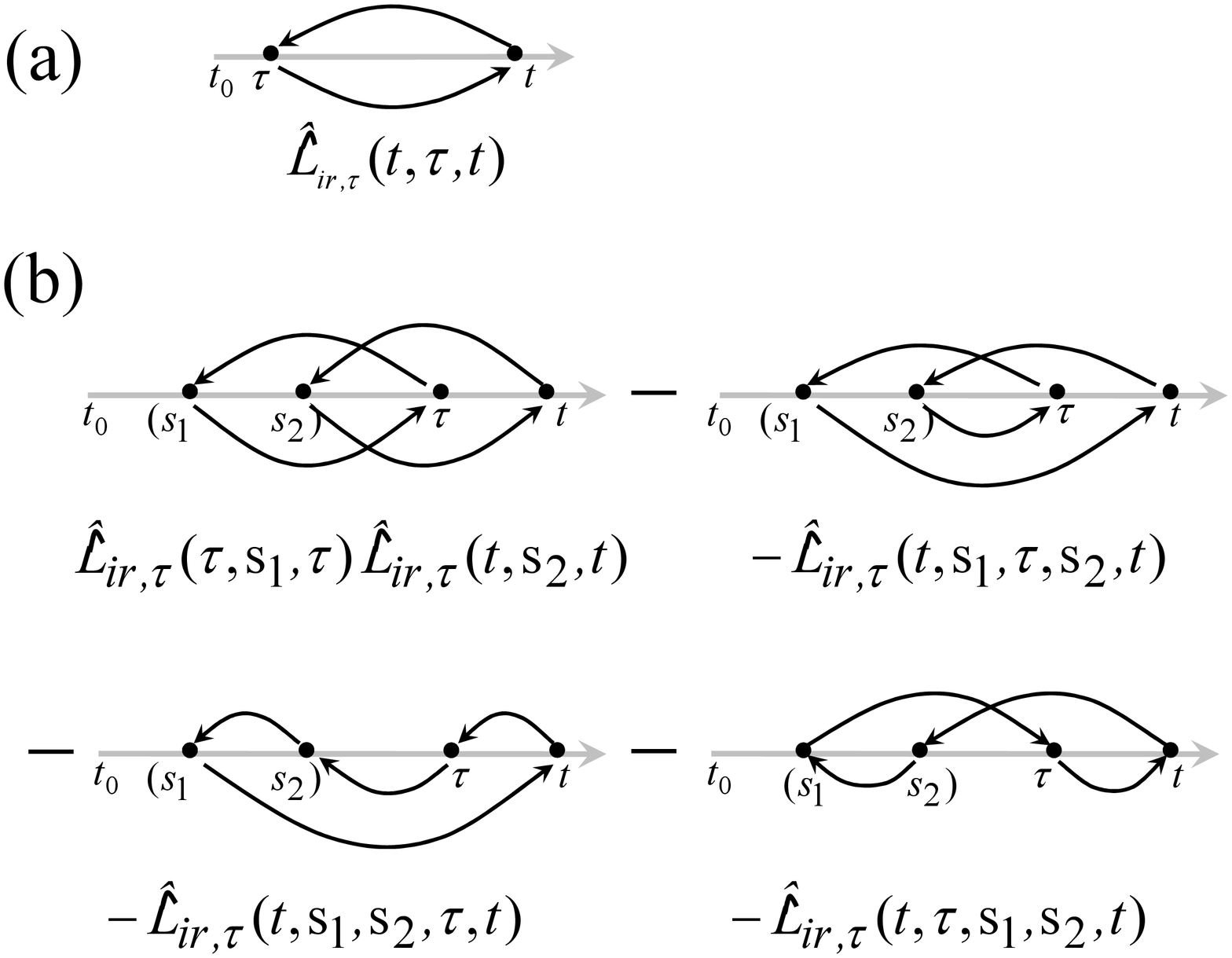}
\caption{Irreducible diagrams and the corresponding loop operators. (a) and
(b) The leading and the second order contributions.}
\label{irker2}
\end{figure}

Meanwhile, each order of the kernel expansion $K^{(n)}(t-\tau )*\rho
_{S}(\tau )$ in Eq. (\ref{eq:ke}) contains multi-particle ($2,4,\cdots ,n+2$
particles) correlations. We can re-express the perturbation expansion of $%
K(t-\tau )*\rho _{S}(\tau )$ in terms of the irreducible diagrams
\begin{eqnarray}
&&K(t-\tau )*\rho _{S}(\tau )  \nonumber \\
&=&K_{ir}^{(0)}(t-\tau )*\rho _{S}(\tau )+K_{ir}^{(2)}(t-\tau )*\rho
_{S}(\tau )+\cdots .  \nonumber \\
&&
\end{eqnarray}
Explicitly, the $n$-th order perturbation $K_{ir}^{(n)}(t-\tau )*\rho
_{S}(\tau )$ contains $n+2$ vertices with the time arguments $\{t,\tau
,s_{1},\cdots ,s_{n}\}$. All vertices are denoted by filled circles without
discriminating $\hat{q}$ and $\hat{q}^{+}$. Also, $n$ internal vertices are
treated indistinguishably, namely, a diagram which exchanges arbitrary two
time arguments of internal vertices is topology-invariant. The
counterclockwise and clockwise loops through the same vertices are also
equivalent except that the vertex orderings along the both loop are
different. Thus, we define \textit{irreducible diagrams }as all \textit{%
connected} topology-independent diagrams\textit{. }The connected diagram
means that each loop of the diagram should intersect with other loops at
least once. As a result, $K_{ir}^{(n)}(t-\tau )*\rho _{S}(\tau )$ comprises
all irreducible diagrams with the prefactor $(-1)^{(n+2)/2+l}$. Each loop of
irreducible diagrams is then given by the irreducible loop operator defined
as follows:
\begin{eqnarray}
&&\mathcal{\hat{L}}_{ir,\tau }\left(
\begin{array}{lllllll}
t & , & s_{1} & ,\cdots \tau \cdots s_{i}\cdots , & s_{2p-2} & , & t
\end{array}
\right)  \nonumber \\
&=&\mathcal{\hat{L}}_{\tau }\left(
\begin{array}{lllllll}
t &  & s_{1} & \cdots \tau \cdots s_{i}\cdots & s_{2p-2} &  & t \\
& l_{1} &  & \qquad \,\cdots &  & r_{p} &
\end{array}
\right)  \nonumber \\
&&+\mathcal{\hat{L}}_{\tau }\left(
\begin{array}{lllllll}
t &  & s_{2p-2} & \cdots s_{i}\cdots \tau \cdots & s_{1} &  & t \\
& r_{p} &  & \qquad \,\cdots &  & l_{1} &
\end{array}
\right) .  \nonumber \\
&&  \label{eq:irexlp}
\end{eqnarray}
The loop operators up to the second order perturbation can be easily
calculated accordingly, see Appendix C. The explicit result of the
irreducible diagrams and the corresponding loop operators is shown in Fig.
\ref{irker2}.

\subsection{Master equation for the reduced density operator}

Thus, writing down the corresponding loop operators one by one according to
the resulted irreducible diagrams, and applying these irreducible loop
operators to the following generating functional
\begin{eqnarray}
&&\mathcal{J}(\hat{q},\hat{q}^{+};\vec{\alpha},\vec{\gamma})=\int dt^{\prime
}\mathcal{C}[\vec{\gamma}(t^{\prime })]\int d\tau ^{\prime }\mathcal{C}[\vec{%
\gamma}^{\prime }(\tau ^{\prime })]  \nonumber \\
&&~~~~~~~~~~~~~~~~~~~~~~\times \mathcal{J}_{in}(\hat{q},\hat{q}^{+};\vec{%
\alpha}),  \nonumber \\
&&~~~~~~~~~~~~  \label{eq:core}
\end{eqnarray}
\begin{eqnarray}
\mathcal{C}[\vec{\gamma}(t)](\cdot ) &=&\Big\{\gamma _{l}(t)\hat{q}%
_{rl}(t)(\cdot )-\gamma _{l}^{*}(t)(\cdot )\hat{q}_{rl}(t)\Big\}  \nonumber
\\
&&+\Big\{\gamma _{r}(t)\hat{q}_{rl}^{+}(t)(\cdot )-\gamma _{r}^{*}(t)(\cdot )%
\hat{q}_{rl}^{+}(t)\Big\},  \label{eq:cl}
\end{eqnarray}
\begin{eqnarray}
&&\mathcal{J}_{in}(\hat{q},\hat{q}^{+};\vec{\alpha})  \nonumber \\
&=&Te^{-i\int ds\{\alpha _{l}(s)\hat{q}_{rl}(s)+\alpha _{r}(s)\hat{q}%
_{rl}^{+}(s)\}}\rho _{S}(t_{0})  \nonumber \\
&&\times \tilde{T}e^{i\int ds^{\prime }\{\alpha _{l}^{*}(s^{\prime })\hat{q}%
_{rl}(s^{\prime })+\alpha _{r}^{*}(s^{\prime })\hat{q}_{rl}^{+}(s^{\prime
})\}},
\end{eqnarray}
and then taking all parameters $(\alpha (s),\gamma (s))$ to be zero, the
explicit expression of $K_{ir}^{(n)}(t-\tau )*\rho _{S}(\tau )$ can be
obtained in terms of coupling operators. Here, the functional derivatives $%
\Big\{\frac{\delta }{\delta \alpha ^{(*)}}\cdots \frac{\delta }{\delta
\gamma ^{(*)}}\Big\}$ in irreducible loop operators are responsible to
generate the correct orderings of the coupling operators, namely, $\{%
\mathcal{S}_{rlr^{\prime }l^{\prime }\cdots r_{n}l_{n};i}\}$.

As a result, we obtain the master equation for the reduced density operator
expressed in terms of the irreducible loop operators to all orders in
perturbation expansions,
\begin{equation}
\frac{\partial }{\partial t}\rho _{S}(t)=-i\left[ H_{S},\rho _{S}(t)\right]
-\int_{t_{0}}^{t}d\tau K(t-\tau )*\rho _{S}(\tau ),  \label{eq:eomirr}
\end{equation}
\begin{eqnarray}
&&K(t-\tau )*\rho _{S}(\tau )=Q_{S}(t-t_{0})  \nonumber \\
&&\times \Big\{\mathcal{\hat{K}}_{ir}(t,\tau )\cdot e^{\mathcal{\hat{W}}%
(\tau )}\cdot \mathcal{J}(\hat{q},\hat{q}^{+};\vec{\alpha},\vec{\gamma})%
\Big\}\Big|_{\vec{\gamma}=\vec{\gamma}^{\prime }=\{\vec{\alpha}\}=0},
\nonumber \\
&&  \label{eq:irreoms}
\end{eqnarray}
where $\mathcal{\hat{K}}_{ir}(t,\tau )$ consists of all allowed irreducible
diagrams (in terms of the irreducible loop operators)
\begin{eqnarray}
&&\mathcal{\hat{K}}_{ir}(t,\tau )=\sum_{n=0}^{\infty }(-1)^{n}\mathcal{\hat{L%
}}_{ir,\tau }\left( t,\mathbf{P}_{\tau }\{\tau ,s_{1},\cdots
,s_{2n}\},t\right)  \nonumber \\
&&+\sum_{n,n^{\prime }=0}^{\infty }(-1)^{n+n^{\prime }}\mathcal{\hat{L}}%
_{ir,\tau }\left( \tau ,s_{1}^{\prime },\cdots ,s_{2n^{\prime }+1}^{\prime
},\tau \right)  \nonumber \\
&&\times \mathcal{\hat{L}}_{ir,\tau }\left( t,s_{1},\cdots ,s_{2n+1},t\right)
\label{eq:irrblek}
\end{eqnarray}
with $\mathbf{P}_{\tau }\{\tau ,s_{1},s_{2},s_{3},\cdots ,s_{n}\}=\{\tau
,s_{1},s_{2},s_{3},\cdots \}+\{s_{1},\tau ,s_{2},s_{3},\cdots \}+\cdots $
containing $n+1$ permutations, and the operator $\mathcal{\hat{W}}(t)$ is
defined as
\begin{equation}
\mathcal{\hat{W}}(t)=\sum_{n=1}^{\infty }\frac{(-1)^{n+1}}{n}\mathcal{\hat{L}%
}_{in,\tau }[s_{1},\cdots s_{2n},s_{1}],  \label{eq:irrblew}
\end{equation}
a log form in the Taylor expansion, and the loop operator $\mathcal{\hat{L}}%
_{in,\tau }[\cdots ]$ only involving the internal vertices is given by
\begin{eqnarray}
&&\mathcal{\hat{L}}_{in,\tau }[
\begin{array}{lll}
s_{1} & ,\cdots s_{2p}, & s_{1}
\end{array}
]  \nonumber \\
&\equiv &\frac{1}{2}\mathcal{\hat{L}}_{ir,\tau }\left(
\begin{array}{lll}
s_{1} & ,\cdots s_{2p}, & s_{1}
\end{array}
\right) .  \nonumber \\
&=&\mathcal{\hat{L}}_{\tau }\left(
\begin{array}{lllllll}
s_{1} &  & s_{2} & \cdots & s_{2p} &  & s_{1}\nonumber \\
& k_{1} &  & \cdots &  & k_{p}^{\prime } &
\end{array}
\right) \\
&=&\mathcal{\hat{L}}_{\tau }\left(
\begin{array}{lllllll}
s_{1} &  & s_{2p} & \cdots & s_{2} &  & s_{1} \\
& k_{p}^{\prime } &  & \cdots &  & k_{1} & \label{eq:irinlp}
\end{array}
\right) .
\end{eqnarray}
In fact, the operator $e^{\mathcal{\hat{W}}(t)}$ generates all the loop
operators for the reduced density operator $\rho _{S}(\tau )$:
\begin{eqnarray}
\rho _{S}(t) &=&Q_{S}(t-t_{0})\Big\{e^{\mathcal{\hat{W}}(t)}\mathcal{J}_{in}(%
\hat{q},\hat{q}^{+};\vec{\alpha})\Big|_{\{\vec{\alpha}\}=0}\Big\}.  \nonumber
\\
&&
\end{eqnarray}

The leading order contribution ($n=0$) to the master equation is obtained as
follows
\begin{eqnarray}
&&\frac{\partial }{\partial t}\rho _{S}(t)=-i\left[ H_{S},\rho
_{S}(t)\right] -Q_{S}(t-t_{0})  \nonumber \\
&&~~~~~~~~~~~~~~~\times \int_{t_{0}}^{t}d\tau K_{ir}^{(0)}[t-\tau ,\rho
_{S}(\tau )],  \label{eq:n0master}
\end{eqnarray}
\begin{eqnarray}
&&K_{ir}^{(0)}[t-\tau ,\rho _{S}(\tau )]=\Big\{\mathcal{\hat{L}}_{ir,\tau
}\left( t,\tau ,t\right) \int dt^{\prime }\mathcal{C}[\vec{\gamma}(t^{\prime
})]  \nonumber  \label{eq:n0kernal} \\
&&~~~~\times \int d\tau ^{\prime }\mathcal{C}[\vec{\gamma}^{\prime }(\tau
^{\prime })]Q_{S}^{-1}(\tau -t_{0})\rho _{S}(\tau )\Big\}\Big|_{\vec{\gamma}=%
\vec{\gamma}^{\prime }=0}.  \nonumber \\
&&
\end{eqnarray}
Also, the time variated reservoir fluctuation due to the interaction with
the qubit has been taken into account. The internal vertices of $\mathcal{%
\hat{K}}_{ir}(t,\tau )$ and $e^{\mathcal{\hat{W}}(\tau )}$ in Eq. (\ref
{eq:irreoms}) are mixed together for higher order contributions. The reduced
density operator can not be extracted unless the following approximation is
utilized,
\begin{eqnarray}
&&\mathcal{\hat{K}}_{ir}(t,\tau )e^{\mathcal{\hat{W}}(\tau )}\mathcal{J}(%
\hat{q},\hat{q}^{+};\vec{\alpha},\vec{\gamma})\Big|_{\vec{\gamma}=\vec{\gamma%
}^{\prime }=\{\vec{\alpha}\}=0}  \nonumber \\
&\approx &\mathcal{\hat{K}}_{ir}(t,\tau )\mathcal{J}_{BA}(\hat{q},\hat{q}%
^{+};\vec{\alpha},\vec{\gamma})\Big|_{\vec{\gamma}=\vec{\gamma}^{\prime }=\{%
\vec{\alpha}\}=0},
\end{eqnarray}
\begin{eqnarray}
&&\mathcal{J}_{BA}(\hat{q},\hat{q}^{+};\vec{\alpha},\vec{\gamma})\equiv \int
dt^{\prime }\mathcal{C}[\vec{\gamma}(t^{\prime })]\int d\tau ^{\prime }%
\mathcal{C}[\vec{\gamma}^{\prime }(\tau ^{\prime })]  \nonumber \\
&&~~~~~~~~~~~~\times Te^{-i\int ds\{\alpha _{l}(s)\hat{q}_{rl}(s)+\alpha
_{r}(s)\hat{q}_{rl}^{+}(s)\}}  \nonumber \\
&&~~~~~~~~~~~~\times \Big\{Q_{S}^{-1}(\tau -t_{0})\rho _{S}(\tau )\Big\}
\nonumber \\
&&~~~~~~~~~~~~\times \tilde{T}e^{i\int ds^{\prime }\{\alpha
_{l}^{*}(s^{\prime })\hat{q}_{rl}(s^{\prime })+\alpha _{r}^{*}(s^{\prime })%
\hat{q}_{rl}^{+}(s^{\prime })\}}.  \nonumber \\
&&
\end{eqnarray}
Without resorting the traditional diagrammatic technique in the Laplace
space, the real-time diagrammatic technique has been developed to derive the
master equation. The charge qubit dynamics in the non-Markovian regime can
be studied based on Eqs. (\ref{eq:eomirr}-\ref{eq:n0kernal}).

\section{QUBIT DECOHERENCE}

To explore the qubit decoherence induced by the QPC measurement, the charge
qubit as a single electron in a double quantum dots\cite{gurvitz} is
considered. The Hamiltonian of the system in Eq. (\ref{eq:sysh}) can be
explicitly written as
\begin{equation}
H_{S}=E_{L}c_{L}^{+}c_{L}+E_{R}c_{R}^{+}c_{R}+\Omega
_{0}(c_{L}^{+}c_{R}+c_{R}^{+}c_{L}),  \label{eq:hqubit}
\end{equation}
where $c_{L}^{+}(c_{L})$ and $c_{R}^{+}(c_{R})$ are the creation
(annihilation) operators of the electron sited in the two dots labeled by $L$
and $R$ with the single-electron constraint $c_{L}^{+}c_{L}+c_{R}^{+}c_{R}=1$%
, $E_{L,R}$ are the corresponding energies, and $\Omega _{0}$ is the
electron hopping amplitude between the double dots. The interaction between
the system and the QPC due to the measurement is characterized by the
interaction coupling\cite{gurvitz}
\begin{equation}
q_{rl}=\Omega (\epsilon _{l},\epsilon _{r})-\Omega ^{\prime }(\epsilon
_{l},\epsilon _{r})c_{R}^{+}c_{R},  \label{eq:tcupling}
\end{equation}
and $\Omega $ and $\Omega -\Omega ^{\prime }$ in Eq.
(\ref{eq:tcupling}) are the electron hoping amplitude of the QPC
without and with the measurement of the single electron in the
double dots. Eq. (\ref{eq:tcupling}) describes a variation in the
barrier of the QPC when the single electron occupies on the right
dot.

We shall consider the leading order contribution to the master equation.
According to Eqs. (\ref{eq:n0master},\ref{eq:n0kernal}), we obtain
\begin{eqnarray}
&&\frac{\partial }{\partial t}\rho _{S}(t)=-i\left[ H_{S},\rho
_{S}(t)\right] -\int_{t_{0}}^{t}d\tau \Big[R_{0},[[k(t-\tau )  \nonumber \\
&&~~~~~\times R(t-\tau ),e^{-iH_{S}(t-\tau )}\rho _{S}(\tau
)e^{iH_{S}(t-\tau )}]]\Big],  \label{eq:kernj}
\end{eqnarray}
where the double bracket $\left[ \left[ A,B\right] \right] =AB-(AB)^{+}$,
the operator $R(t)$ is given by
\begin{eqnarray}
R(t) &=&\cos \theta (\left| e\right\rangle \left\langle e\right| -\left|
g\right\rangle \left\langle g\right| )  \nonumber \\
&&-\sin \theta (e^{i\gamma t}\left| g\right\rangle \left\langle e\right|
+e^{-i\gamma t}\left| e\right\rangle \left\langle g\right| )  \label{eq:r}
\end{eqnarray}
with $\gamma =\sqrt{4\Omega _{0}^{2}+(E_{L}-E_{R})^{2}}$ the energy
difference between the ground state $\left| g\right\rangle $ and the
excited state $\left| e\right\rangle $ of the qubit, $\theta =\cos
^{-1}[(E_{L}-E_{R})/\gamma ]$, and $R_{0}=R(t=0)$. The reservoir
correlation function $k(s)$ in Eqs. (\ref{eq:kernj}) which
characterizes the QPC structure associated with the temperature
effect and the external bias is expressed as
\begin{equation}
k(s)=\int_{-\infty }^{\infty }\frac{d\epsilon d\epsilon ^{\prime }}{4\pi ^{2}%
}e^{i(\epsilon -\epsilon ^{\prime })s}\tilde{k}(\epsilon ,\epsilon ^{\prime
}),  \label{eq:k}
\end{equation}
\begin{eqnarray}
\tilde{k}(\epsilon ,\epsilon ^{\prime }) &=&f_{R}(\epsilon
)(1-f_{L}(\epsilon ^{\prime }))J(\epsilon ,\epsilon ^{\prime })
\label{eq:ksptrum} \\
&&+f_{L}(\epsilon )(1-f_{R}(\epsilon ^{\prime }))J(\epsilon ^{\prime
},\epsilon ),  \nonumber
\end{eqnarray}
where $\tilde{k}(\epsilon ,\epsilon ^{\prime })$ is the electron-tunneling
spectrum for the QPC, $f_{L,R}(\epsilon )=1/(1+\exp \beta (\epsilon -\mu
_{L,R}))$ are respectively the Fermi-Dirac distribution functions for the
source and the drain, and the spectral density $J(\epsilon ^{\prime
},\epsilon )$ for the QPC structure is given by
\begin{equation}
J(\epsilon ,\epsilon ^{\prime })=\pi ^{2}g_{L}(\epsilon ^{\prime
})g_{R}(\epsilon )\left| \Omega ^{\prime }(\epsilon ,\epsilon ^{\prime
})\right| ^{2}  \label{eq:j2}
\end{equation}
with $g_{L,R}(\epsilon )$ being the density of states of the source and the
drain. Also, we assume here the electron energy levels $\{\epsilon
_{l},\epsilon _{r}\}$ are continuous. Eq. (\ref{eq:kernj}) contains the
non-Markovian processes of the qubit dynamics up to the leading order. It
can be checked Eq. (\ref{eq:kernj}) reduces to the result in Refs. \cite
{li,lee} in the Markovian limit.

\begin{widetext}

\begin{figure}[tbph]
\centering
\includegraphics*[angle=270,scale=.4]{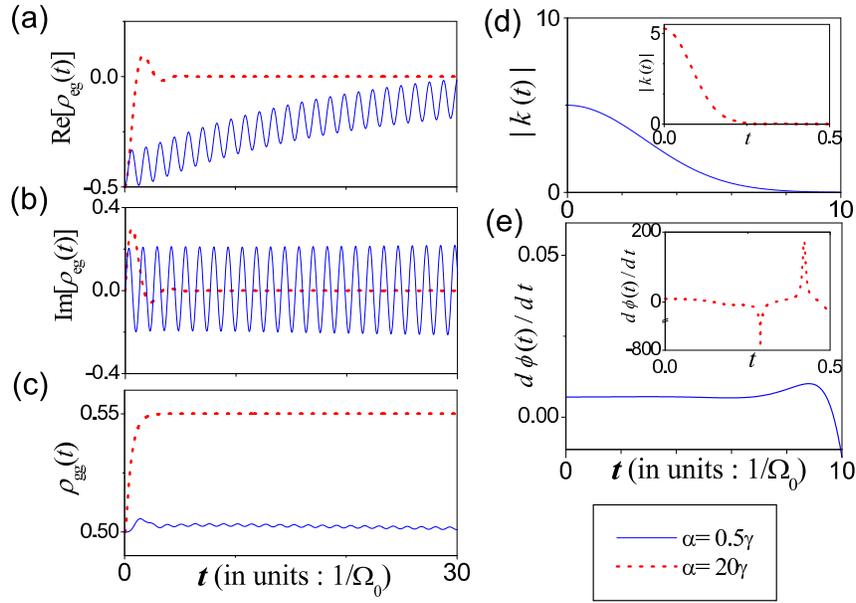}
\caption{(color online). The qubit dynamics and the corresponding
reservoir time correlation function in the measurement condition
with an extreme small $w$ ($w\ll\gamma$) and different $\alpha$. The
symmetric coupled quantum dots ($\gamma =2\Omega _{0}$) is
simulated. The qubit is set initially in the $|L\rangle $ state. The
following measurement parameters are used with $\Omega _{0}$ being a
rescaling factor: $\beta =1/\Omega _{0}$, $V_{d}=20\Omega _{0}$ and
$p=\Omega _{0}$. (a) and (b) The qubit dephasing. The time evolution
of the real and imaginary parts of the qubit density matrix element
$\rho _{eg}$ are plotted, respectively. (c) The qubit relaxation.
The time evolution of the qubit density matrix element $\rho _{gg}$
for the qubit in the ground state is plotted. (d) The amplitude of
the reservoir time correlation function is plotted in units $\Omega
_{0}^{2}$. (e) The time derivative of the phase in the reservoir
time correlation function is plotted in units $\Omega _{0}$. The
integration in Eq. (\ref{eq:k2a}) is approximately calculated by
fitting the Lorentzian spectral density with Gaussian function to
remove the energy cut-off. } \label{w_zero}
\end{figure}
\end{widetext}

To make the non-Markovian feature apparent, we shall concentrate on the
charge qubit with symmetric coupled dots $E_{L}=E_{R}$ characterized by $%
\gamma =2\Omega _{0}$ ($\theta =\pi /2$). The equation of motion for the
reduced density matrix becomes,
\begin{eqnarray}
\frac{\partial \rho _{eg}(t)}{\partial t} &=&-i\gamma \rho
_{eg}(t)-2\int_{t_{0}}^{t}d\tau |k(t-\tau )|\cos \phi (t-\tau )  \nonumber \\
&&\times \Big\{\rho _{eg}(\tau )-\rho _{ge}(\tau )\Big\},  \label{eq:eg}
\end{eqnarray}
\begin{eqnarray}
\frac{\partial \rho _{gg}(t)}{\partial t} &=&2\int_{t_{0}}^{t}d\tau
|k(t-\tau )|  \nonumber \\
&&\times \Big\{\cos [(t-\tau )\gamma -\phi (t-\tau )]\rho _{ee}(\tau )
\nonumber \\
&&-\cos [(t-\tau )\gamma +\phi (t-\tau )]\rho _{gg}(\tau )\Big\},
\label{eq:gg}
\end{eqnarray}
where the matrix elements are defines as $\rho _{ij}(t)=\left\langle
i|\rho _{S}(t)|j\right\rangle $ with $\left| i,j\right\rangle $
being the ground state or the excited state of the qubit, and $\phi
(t)$ is the phase of the reservoir correlation function
$k(t)=|k(t)|e^{-i\phi (t)}$.

The qubit decoherence can be studied by the analysis of the spectral
density $J(\epsilon _{r},\epsilon _{l})$. In the Literature, the
density of states in the QPC reservoirs and the hopping amplitude
across the QPC barrier are assumed to be energy-level independent,
namely, the wide-band approximation for the QPC structure.\cite
{gurvitz,korotkov,goan,shnirman,stace,li,lee} The spectral density
is then given by
\begin{equation}
J(\epsilon _{r},\epsilon _{l})=\pi ^{2}g_{L}g_{R}\left| \Omega ^{\prime
}\right| ^{2}.  \label{eq:tpflat}
\end{equation}
This corresponds to the Markovian limit, in which the qubit dynamics
is strongly decoherent\cite{goan,stace,li,lee}. The Markovian
dynamics arises from tunneling electrons in the QPC with a shortest
time correlation $k(t-\tau )\propto \delta (t-\tau )$.

However, as indicated in Eqs. (\ref{eq:k},\ref{eq:ksptrum}), the QPC
structure determines the correlation time scale of the tunneling
electron fluctuation in the QPC. The non-Markovian processes of the
qubit dynamics emerges only when the QPC structure is casted with a
finite correlation time scale. The effect of the QPC structure can
be characterized by an energy-level dependence of the spectral
density. In the literature, the spectral density $J_{RD}(\epsilon )$
for a reservoir coupling to a dot (or a molecular wire) is
parameterized by Lorentzian spectrums,\cite{welack}
\begin{equation}
J_{RD}(\epsilon )=\frac{\pi }{2}\sum_{k=1}^{m}\frac{p_{k}\alpha _{k}}{%
(\epsilon -w_{k})^{2}+\alpha _{k}^{2}/4},
\end{equation}
where $p_{k}$, $w_{k}$ and $\alpha _{k}$ are fitting parameters. The
Lorentzian spectrum has also been applied to study the quantum measurement
of this system (using a constant hopping amplitude with a Lorentzian density
of state).\cite{spectrum2} For our system which involves electrons tunneling
a barrier between two reservoirs, the spectral density can be approximately
treated as
\begin{equation}
J(\epsilon _{r},\epsilon _{l})=\frac{\pi }{2}\frac{p\alpha }{(|\epsilon
_{r}-\epsilon _{l}|-w)^{2}+\alpha ^{2}/4}.  \label{eq:resolutf}
\end{equation}
In Eq. (\ref{eq:resolutf}), the parameter $p$ specifies the magnitude of the
spectral density, $w$ characterizes the variation of the barrier potential
due to the interaction with the qubit electron, and the width $\alpha $
indicates a modulation of the decay rate for the qubit decoherence. Changing
the variables of the integration in Eq. (\ref{eq:k}) from $(\epsilon
,\epsilon ^{\prime })$ to $(\varepsilon ^{\prime }=\frac{\epsilon +\epsilon
^{\prime }}{2},\varepsilon =\frac{\epsilon ^{\prime }-\epsilon }{2})$ and
integrating out $\varepsilon ^{\prime }$, we then obtain the following
reservoir correlation function,
\begin{eqnarray}
k(s) &=&\int_{-\infty }^{\infty }\frac{d\varepsilon }{2\pi }%
e^{-is\varepsilon }\frac{p\alpha }{4(|\varepsilon |-w)^{2}+\alpha ^{2}}
\nonumber \\
&&\qquad \times \Big\{\tilde{g}(\varepsilon -V_{d})+\tilde{g}(\varepsilon
+V_{d})\Big\},  \label{eq:k2a}
\end{eqnarray}
where the function $\tilde{g}(x)=\frac{x}{1-e^{-\beta x}}$.

\begin{figure}[tbph]
\includegraphics*[angle=270,scale=.5]{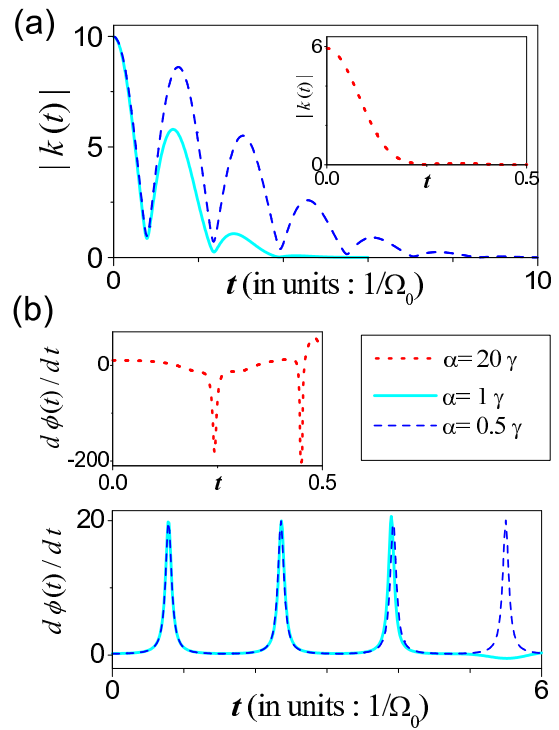}
\caption{(color online). The reservoir time correlation function
plotted in the measurement condition that the variation of the QPC
barrier potential has the same energy scale as the qubit system
($w=\gamma$) and with different $\alpha$. The symmetric coupled
quantum dots ($\gamma =2\Omega _{0}$) is considered. The following
measurement parameters are used with $\Omega _{0}$ being a rescaling
factor: $\beta =1/\Omega _{0}$, $V_{d}=20\Omega _{0}$ and $p=\Omega
_{0}$, respectively. (a) The amplitude of the reservoir time
correlation function is plotted in units $\Omega _{0}^{2}$. (b) The
time derivative of the phase
of the reservoir time correlation function is plotted in units $\Omega _{0}$%
. }
\label{k2}
\end{figure}

A close connection between the qubit decoherence and the tunneling-electron
fluctuation is revealed in Eq. (\ref{eq:k2a}). According to the spectral
density in Eq. (\ref{eq:resolutf}), the randomness of tunneling electrons
across the QPC barrier is expounded first. Obviously, taking the limit $%
\alpha \ll \left| \epsilon _{l}-\epsilon _{r}\right| $ to Eq. (\ref
{eq:resolutf}), $J(\epsilon _{r},\epsilon _{l})$ reduces to $\pi
^{2}\delta (\left| \epsilon _{l}-\epsilon _{r}\right| -w)$, only two
channels $\epsilon _{l}-\epsilon _{r}=\pm w$ involved. On the other
hand, $\alpha \gg \left| \epsilon _{l}-\epsilon _{r}\right| $ leads
to a channel-mixture regime, where all transitions $\epsilon
_{l}\rightleftarrows \epsilon _{r}$ that electrons tunneling between
the source and the drain are allowed with the weight determined by
$J(\epsilon _{r},\epsilon _{l})$. The randomness of electron
tunneling processes in the channel-mixture regime comes from
electron scattering which are determined by the band structure
associated with the geometry of the two metal gates in the QPC and
modulated by the interaction with the qubit electron. The parameter
$\alpha $ in Eq. (\ref
{eq:resolutf}) characterizes the deviation describing how transfer energies $%
\left| \epsilon _{l}-\epsilon _{r}\right| $ in all electron tunneling
processes are close to $w$ in the statistics of ensemble average. The qubit
dynamics with different $\alpha $ and an extreme small $w(\ll \gamma )$ are
simulated in Fig. \ref{w_zero} with (a) and (b) for the qubit dephasing and
(c) the qubit relaxation. The result shows that the qubit decoherence is
suppressed as tunneling electrons with a smaller deviation. Associated with
the qubit decay rate, $\alpha $ can be used to judge the decoherent behavior
of the qubit state.

\begin{widetext}

\begin{figure}[tbph]
\centering
\includegraphics*[angle=270,scale=.4]{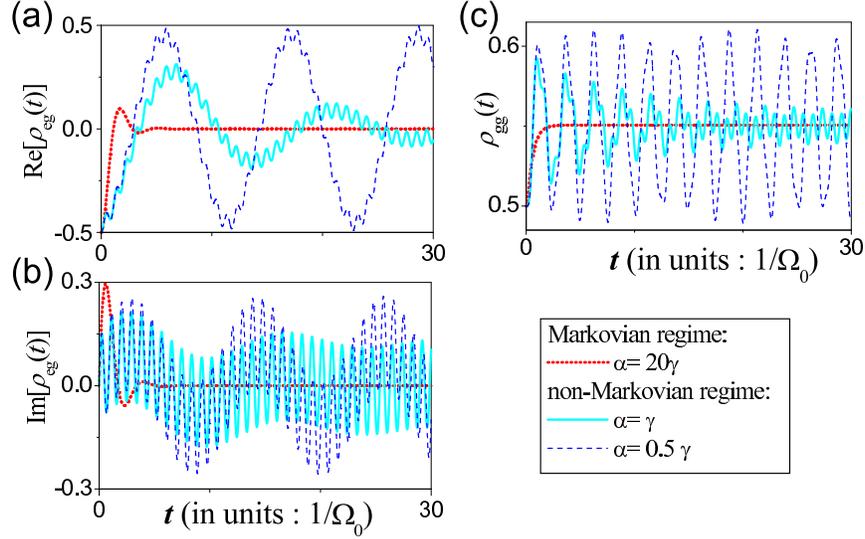}
\caption{(color online). The qubit dynamics measured in the
condition $w=\gamma$ and with different $\alpha$. The symmetric
coupled quantum dots ($\gamma =2\Omega _{0}$) is simulated. The
qubit is set initially in the $|L\rangle $ state. The following
measurement parameters are used with $\Omega _{0}$ being a rescaling
factor: $\beta =1/\Omega _{0}$, $V_{d}=20\Omega _{0}$ and $p=\Omega
_{0}$, respectively. (a) and (b) for the qubit dephasing, and (c)
the qubit relaxation.} \label{qdynamics}
\end{figure}
\end{widetext}

Furthermore, the reservoir time correlation function in Eq. (\ref{eq:k2a})
describes the time correlation of tunneling-electron fluctuation in the
variation due to the measurement. In Fig. \ref{k2}, the reservoir time
correlation function is plotted in the condition that the variation of the
QPC barrier potential has the same energy scale as the qubit system. It can
be found in Fig. \ref{k2} (a) that the amplitude $|k(s)|$ is a Lorentzian
profile with periodic deep peaks at $t=\pi /2w,$ $3\pi /2w,$ $\cdots $. The
amplitude $|k(t)|$ describes the correlation time scale between tunneling
electrons which are induced by the measurement in the time interval $t$.
With fixing the amplitude, the wider the half width of the Lorentzian
profile of $|k(t)|$, the longer the time correlation of tunneling-electron
fluctuation. Fig. \ref{k2} (a) then indicates that the time correlation
increases with $\alpha $ decreasing. In other words, a shorter time
correlation between electron tunneling processes leads to a more random
electron-tunneling spectrum (with wide profile), see Eqs. (\ref{eq:k},\ref
{eq:ksptrum}). In the channel-mixture regime, the time correlation of
tunneling-electron fluctuation is smeared, and the reservoir memory effect
on the qubit dynamics is suppressed. The qubit dynamics with different $%
\alpha $ are simulated in Fig. \ref{qdynamics}. The result shows
that in the channel-mixture regime ($\alpha =20\gamma $), the qubit
undergoes a severe decoherence, which corresponds to the qubit
dynamics in the Markovian limit. On the other hand, instructively,
if the QPC structure can be recasted such that the measurement
approaches to the case almost without random electron scatterings
($\alpha =0.5\gamma $), the non-Markovian processes of the qubit
dynamics emerges. The qubit simply performs a periodic oscillation
with small fluctuations. The qubit decoherence is suppressed. The
mechanism of the qubit decoherence can be understood by the analysis
of the time derivative of the phase $\phi (t)$ in the reservoir time
correlation function, which is plotted in Fig. \ref{k2} (b).
According to Eq. (\ref {eq:gg}), $d\phi (t)/dt$ characterizes the
average transfer energy that the qubit electron obtains from all
tunneling processes in the QPC. Fig. \ref{k2} (b) indicates that in
the channel-mixture regime ($\alpha =20\gamma $), the negative
energy-transfer rate before $t=0.25/\Omega _{0}$ leads to the qubit
electron energy being exhausted. The contribution that the qubit
electron absorbs energy from the QPC is very small and ignorable
after $t=0.25/\Omega _{0}$, because $|k(t)|$ decay to zero, see Fig.
\ref{k2} (a). This effect occurs repeatedly due to the non-Markovian
dynamical structure. The qubit finally relaxes to a mixed state, as
shown in Fig. \ref{qdynamics}. However, for the case $\alpha
=0.5\gamma $, the qubit electron absorbs (emits) energy from (to)
the QPC before (after) $t=\pi /2w,$ $3\pi /2w,$ $\cdots $ in each
period, as shown by the peaks with positive (negative)
energy-transfer rates in Fig. \ref{k2} (b). The qubit is driven by
the QPC periodically. Because of a longer time correlation of the
tunneling-electron fluctuation [see the wide profile of $|k(t)|$ in
Fig. \ref{k2} (a)], this periodic driving becomes significant. The
qubit state is oscillating almost without decoherence, if the effect
of the random electron scatterings can be ignored.

Accordingly, the qubit decoherence also depends on the number of the
peaks, which is determined by the variation of the barrier
potential. By comparing Fig. \ref{w_zero} (d) and (e) with Fig.
\ref{k2}, it can be found that the number of the periodic peaks
inside the half-width of the reservoir time correlation function
indeed decreases with the parameter $w$ decreasing. The more the
sharp peaks with positive energy-transfer rates that the reservoir
time correlation function contains, the larger amplitude of the
oscillation the qubit performs in the asymptotic regime, specially,
the population $\rho _{gg}(t)$ and the coherence Re$[\rho
_{eg}(t)]$. A large amount of energy transferred between the qubit
electron and the tunneling electrons in the case with large $w$
leads to a severer time variation (oscillation) on the
qubit-electron population and the induced coherence. However, the $w$%
-dependence of the qubit dynamics is not so sensitive for the large $\alpha $%
. The qubit decoherence under the QPC measurement with different $w$ almost
coincide, as indicated by the case of $\alpha =20\gamma $ in Fig. \ref
{w_zero} and Fig. \ref{qdynamics}. To drive the qubit state efficiently,
tunneling electrons must be in the state with a longer time correlation, no
matter how large the variation of the barrier is. Also, Fig. \ref{w_zero}
(b) demonstrates that a sinusoidal quantum oscillation of the qubit
coherence is still accessible under the quantum measurement, if the QPC
structure is casted with a small $w$ of the spectral density.

Finally, the numerical result also shows that the larger the bias voltage is
applied, the shorter the time correlation of the tunneling-electron
fluctuation becomes. A large amount of electrons tunneling across the QPC
barrier forced by the large bias voltage leads to the time correlation of
the tunneling-electron fluctuation being smeared with a sharp profile $%
|k(t)| $. Therefore, the qubit dynamics becomes typically Markovian, and the
qubit state is severely decoherent. The same decoherence behavior also
occurs at a higher temperature. As a result, the constant spectral density
used in literatures essentially describes electron tunneling processes with
the most random channel mixture, and the qubit simply undergoes decoherent
processes. More realistic non-Markovian dynamics emerges only when the
effect of the explicit spectral density is taken into account. A particular
design to cast the QPC structure such that tunneling electrons are in the
state with longer time correlation could suppress the qubit decoherence
during the QPC measurement.

\section{SUMMARY}

We have developed a non-equilibrium theory for the charge qubit
dynamics accompanied with the QPC measurement. The effect of
non-equilibrium fluctuation of the QPC reservoirs to the qubit
dynamics has been treated by using the real-time diagrammatic
technique developed based on Schwinger-Keldysh's approach. We
introduce the loop operator to exactly derive the master equation
for the reduced density operator, which is expressed in terms of the
irreducible diagrams. The qubit decoherence is studied according to
the resulted master equation up to the leading order. The
non-Markovian processes in the qubit dynamics has been taken into
account in this framework. We find that the qubit decoherence
sensitively depends on the spectral density of the QPC structure.
The constant spectral density in fact describes the electron
tunneling processes with the most random channel mixture, which
causes the severest qubit decoherence, and corresponds to the qubit
dynamics in the Markovian limit. However, if the QPC structure can
be controlled such that the spectral density with a narrower width,
the electron tunneling processes reduce to a less channel mixture.
The non-Markovian processes of the qubit dynamics emerges. The qubit
simply performs a periodic oscillation with less decoherence. A
longer time correlation of the tunneling-electron fluctuation
results in the qubit state being periodically driven. In the
channel-mixture regime, the time correlation of the
tunneling-electron fluctuation is smeared by random electron
scatterings. The qubit fails to be driven effectively by the QPC due
to short time correlations of the tunneling-electron fluctuation,
and the qubit state is simply decoherent. As the measurement
operation is designed with the minimum deviation of electron
tunneling processes, the qubit state could be measured with the
least decoherence effect. The further work taking into account
higher order contributions to the qubit decoherence and the noise
spectrum of the QPC output signal is in progress.

\acknowledgements

The authors would like to thank S. A. Gurvitz for useful discussions. The
work is supported by the National Center for Theoretical Science and
National Cheng Kung University of Republic of China under Contract No. OUA
96-3-2-085, and National Science Council of Republic of China under Contract
Nos. NSC-95-2112-M-006-001 and NSC-96-2112-M-006-011-MY3.

\appendix

\section{DERIVATION OF EQUATION OF MOTION IN EQUATION (\ref{eq:eoms})}

The equation of motion for the reduced density operator listed in Eq. (\ref
{eq:eoms}) is derived in this appendix. In the interaction picture, the
formal solution of the Liouville equation $\frac{\partial }{\partial t}\rho
_{tot}=-i\left[ H,\rho _{tot}\right] $ for the total density operator is
given by
\begin{equation}
\hat{\rho}_{tot}(t)=\hat{\rho}_{tot}(t_{0})-i\int_{t_{0}}^{t}d\tau [\hat{H}%
_{\tau }^{\prime },\hat{\rho}_{tot}(\tau )],  \label{eq:eom1}
\end{equation}
iteratively, which leads to Dyson's perturbation expansion for the total
density operator. However, to systematically build up the perturbation
scheme for the kernel expansion in terms of the reservoir contour-order
Green's functions (two-point correlation functions), alternatively, the
second iteration of Eq. (\ref{eq:eom1})
\begin{eqnarray}
&&\hat{\rho}_{tot}(t)=\hat{\rho}_{tot}(t_{0})-i\int_{t_{0}}^{t}d\tau [\hat{H}%
_{\tau }^{\prime },\hat{\rho}_{tot}(t_{0})]  \nonumber \\
&&~~~~~-\int_{t_{0}}^{t}d\tau \int_{t_{0}}^{\tau }d\tau ^{\prime }[\hat{H}%
_{\tau }^{\prime },[\hat{H}_{\tau ^{\prime }}^{\prime },\hat{\rho}%
_{tot}(\tau ^{\prime })]],   \label{eq:eom2}
\end{eqnarray}
which contains Hamiltonian operator orderings of $\hat{H}_{\tau }^{\prime }$
and $\hat{H}_{\tau ^{\prime }}^{\prime }$ in different time should be used.
Obviously, as indicated by Eq. (\ref{eq:eom2}), the total density operator
obeys the following equation of motion
\begin{equation}
\frac{\partial }{\partial t}\hat{\rho}_{tot}(t)=-i\left[ \hat{H}_{t}^{\prime
},\hat{\rho}_{tot}(t_{0})\right] -\int_{t_{0}}^{t}d\tau \left[ \hat{H}%
_{t}^{\prime },\left[ \hat{H}_{\tau }^{\prime },\hat{\rho}_{tot}(\tau
)\right] \right] .
\end{equation}
Then, taking the partial trace to integrate out the degrees of freedom of
the QPC reservoirs, Eq. (\ref{eq:eoms}) is resulted.

\section{EXAMPLE OF CALCULATING THE COEFFICIENT $\mathcal{S}_{rl\cdots ;i}$}

\begin{figure}[tbph]
\includegraphics*[angle=270,scale=.35]{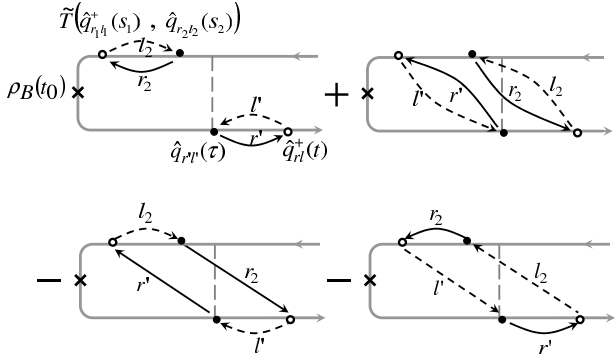}
\caption{Four topology-independent diagrams of the example Eq. (\ref{eq:ex2o}%
).}
\label{2order}
\end{figure}

In this appendix, we present an example of the contribution of the
perturbation expansion Eq. (\ref{eq:pe}) for $n=2$ with the particular
contour ordering $\{t\geq _{p}\tau \geq _{p}t_{0}\geq _{p}(s_{1},s_{2})\}$
to illustrate how to diagrammatically calculate the coefficient $\mathcal{E}%
_{rlr^{\prime }l^{\prime }r_{1}l_{1}r_{2}l_{2}}$ in Eq. (\ref{eq:ptrcoef}),
\begin{eqnarray}
&&\mathcal{S}_{rlr^{\prime }l^{\prime }r_{1}l_{1}r_{2}l_{2}}\Big(t\geq
_{p}\tau \geq _{p}t_{0}\geq _{p}(s_{1},s_{2})\Big)  \nonumber \\
&&\times \mathcal{E}_{rlr^{\prime }l^{\prime }r_{1}l_{1}r_{2}l_{2}}\Big(%
t\geq _{p}\tau \geq _{p}t_{0}\geq _{p}(s_{1},s_{2})\Big)  \nonumber \\
&=&\hat{q}_{rl}^{+}(t)\hat{q}_{r^{\prime }l^{\prime }}(\tau )\rho _{S}(t_{0})%
\tilde{T}\Big\{\hat{q}_{r_{1}l_{1}}^{+}(s_{1})\hat{q}_{r_{2}l_{2}}(s_{2})%
\Big\}  \nonumber \\
&&\times \text{Tr}_{B}\Big[\Big(\hat{\psi}_{l}^{+}(t)\hat{\psi}_{r}(t)\Big)%
\Big(\hat{\psi}_{r^{\prime }}^{+}(\tau )\hat{\psi}_{l^{\prime }}(\tau )\Big)%
\rho _{B}^{(0)}  \nonumber \\
&&\times \tilde{T}\Big\{\Big(\hat{\psi}_{l_{1}}^{+}(s_{1})\hat{\psi}%
_{r_{1}}(s_{1})\Big)\Big(\hat{\psi}_{r_{2}}^{+}(s_{2})\hat{\psi}%
_{l_{2}}(s_{2})\Big)\Big\}\Big].  \label{eq:ex2o}
\end{eqnarray}
The corresponding diagrams are shown in Fig.~\ref{2order}.
It includes four topology-independent diagrams, two single-loop diagrams
with the prefactor $(-1)^{(2+2)/2+1}=-1$ and two double-loop diagrams with
the prefactor $(-1)^{(2+2)/2+2}=1$. The explicit expression of the
coefficient $\mathcal{E}_{rlr^{\prime }l^{\prime }r_{1}l_{1}r_{2}l_{2}}$ of
Eq. (\ref{eq:ex2o}) that the partial trace is carried out can be directly
written down from four topology-independent diagrams
\begin{eqnarray}
&&\mathcal{E}_{rlr^{\prime }l^{\prime }r_{1}l_{1}r_{2}l_{2}}\Big(t\geq
_{p}\tau \geq _{p}t_{0}\geq _{p}(s_{1},s_{2})\Big)  \nonumber \\
&=&\delta _{l_{1}l_{2}}\delta _{r_{1}r_{2}}\delta _{rr^{\prime }}\delta
_{ll^{\prime }}G_{\tilde{F},l_{2}}^{(0)}(s_{2},s_{1})G_{\tilde{F}%
,r_{2}}^{(0)}(s_{1},s_{2})  \nonumber \\
&&\times G_{F,r^{\prime }}^{(0)}(t,\tau )G_{F,l^{\prime }}^{(0)}(\tau ,t)
\nonumber \\
&&+\delta _{l_{1}l^{\prime }}\delta _{r_{1}r^{\prime }}\delta
_{l_{2}l}\delta _{r_{2}r}G_{<,l^{\prime }}^{(0)}(\tau ,s_{1})G_{>,r\prime
}^{(0)}(s_{1},\tau )  \nonumber \\
&&\times G_{>,l_{2}}^{(0)}(s_{2},t)G_{<,r_{2}}^{(0)}(t,s_{2})  \nonumber \\
&&-\delta _{l_{1}l_{2}}\delta _{r_{2}r}\delta _{ll^{\prime }}\delta
_{r_{1}r^{\prime }}G_{\tilde{F}%
,l_{2}}^{(0)}(s_{2},s_{1})G_{<,r_{2}}^{(0)}(t,s_{2})  \nonumber \\
&&\times G_{F,l^{\prime }}^{(0)}(\tau ,t)G_{>,r\prime }^{(0)}(s_{1},\tau )
\nonumber \\
&&-\delta _{l_{1}l^{\prime }}\delta _{rr^{\prime }}\delta
_{r_{1}r_{2}}\delta _{l_{2}l}G_{<,l^{\prime }}^{(0)}(\tau
,s_{1})G_{F,r^{\prime }}^{(0)}(t,\tau )  \nonumber \\
&&\times G_{>,l_{2}}^{(0)}(s_{2},t)G_{\tilde{F},r_{2}}^{(0)}(s_{1},s_{2}).
\end{eqnarray}
All coefficients $\mathcal{E}_{rlr^{\prime }l^{\prime }\cdots }$ for higher
order contributions with arbitrary contour ordering can be calculated
diagrammatically in the similar way.

\section{THE LOOP OPERATORS FOR LEADING AND SECOND ORDER PERTURBATIONS}

\begin{figure}[tbph]
\includegraphics*[angle=270,scale=.36]{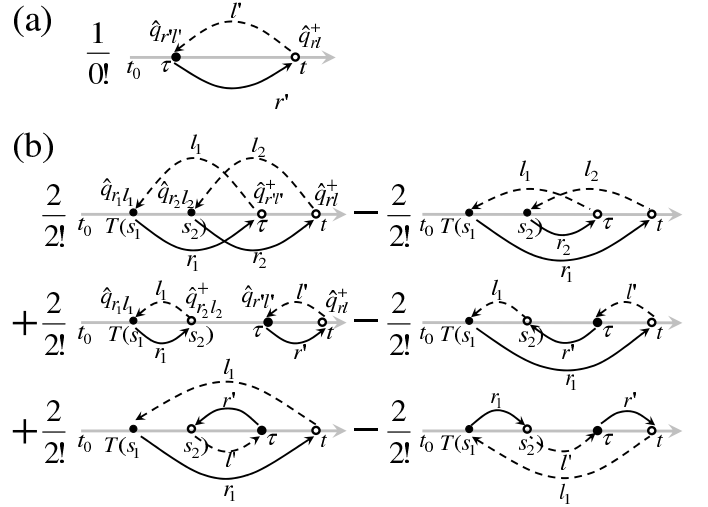}
\caption{Topology-independent diagrams of $K(t-\tau )*\rho _{S}(\tau )$ up
to the second order perturbation. (a) and (b) The leading and the second
order contributions}
\label{ker2}
\end{figure}

As an example, we illustrate how to write down the loop operators for the
leading order kernel $K^{(0)}(t-\tau )*\rho _{S}(\tau )$ and the second
order kernel $K^{(2)}(t-\tau )*\rho _{S}(\tau )$ diagrammatically. For the
leading order kernel, only two topology-independent diagrams involves, and
one is dual to the other. One of the both is shown in Fig.~\ref{ker2} (a).

For the second order kernel with all possible time contour orderings, it
contains totally 24 diagrams, half of them are dual to each other. We only
need to consider one set of them ($12$ diagrams). Furthermore, one can find
that only six topology-independent diagrams exist, see Fig.~\ref{ker2} (b).
Each one corresponds to two topology-equivalent diagrams, i.e. the weight
factor is 2 for each topology-independent diagram, while the prefactor is $%
\frac{-1}{2!}$ for the single-loop diagrams and $\frac{+1}{2!}$ for the
double-loop diagrams.

The loop operators can be directly written down according to the
topology-independent diagrams in Fig.~\ref{ker2}. For the leading order
kernel, the corresponding loop operator is given by $\frac{1}{0!}\mathcal{%
\hat{L}}_{\tau }\left(
\begin{array}{lllll}
t &  & \tau &  & t \\
& r^{\prime } &  & l^{\prime } &
\end{array}
\right) .$ We thus obtain the following expression
\begin{eqnarray}
&&K^{(0)}(t-\tau )*\rho _{S}(\tau )  \nonumber \\
&=&Q_{S}(t-t_{0})\Big\{\mathcal{\hat{L}}_{\tau }\left(
\begin{array}{lllll}
t &  & \tau &  & t \\
& r^{\prime } &  & l^{\prime } &
\end{array}
\right)  \nonumber \\
&&\times \mathcal{J}(\hat{q},\hat{q}^{+};\vec{\alpha},\vec{\gamma})\Big|_{%
\vec{\gamma}=\vec{\gamma}^{\prime }=\{\vec{\alpha}\}=0}+\text{duality}\Big\}.
\label{eq:loopk0}
\end{eqnarray}
Similarly, the explicit expression of the second order kernel can also be
obtained according to the topology-independent diagrams in Fig.~\ref{ker2}
(b),
\begin{eqnarray}
&&K^{(2)}(t-\tau )*\rho _{S}(\tau )  \nonumber \\
&=&Q_{S}(t-t_{0})\Big\{\mathcal{\hat{K}}^{(2)}(t-\tau )\mathcal{J}(\hat{q},%
\hat{q}^{+};\vec{\alpha},\vec{\gamma})\Big|_{\vec{\gamma}=\vec{\gamma}%
^{\prime }=\{\vec{\alpha}\}=0}  \nonumber \\
&&+\text{duality}\Big\},
\end{eqnarray}
\begin{eqnarray}
&&\mathcal{\hat{K}}^{(2)}(t-\tau )  \nonumber \\
&=&\frac{2}{2!}\mathcal{\hat{L}}_{\tau }\left(
\begin{array}{lllll}
\tau &  & s_{1} &  & \tau \\
& r_{1} &  & l_{1} &
\end{array}
\right) \cdot \mathcal{\hat{L}}_{\tau }\left(
\begin{array}{lllll}
t &  & s_{2} &  & t \\
& r_{2} &  & l_{2} &
\end{array}
\right)  \nonumber \\
&&-\frac{2}{2!}\mathcal{\hat{L}}_{\tau }\left(
\begin{array}{lllllllll}
t &  & s_{1} &  & \tau &  & s_{2} &  & t \\
& r_{1} &  & l_{1} &  & r_{2} &  & l_{2} &
\end{array}
\right)  \nonumber \\
&&+\frac{2}{2!}\mathcal{\hat{L}}_{\tau }\left(
\begin{array}{lllll}
s_{2} &  & s_{1} &  & s_{2} \\
& r_{1} &  & l_{1} &
\end{array}
\right) \cdot \mathcal{\hat{L}}_{\tau }\left(
\begin{array}{lllll}
t &  & \tau &  & t \\
& r^{\prime } &  & l^{\prime } &
\end{array}
\right)  \nonumber \\
&&-\frac{2}{2!}\mathcal{\hat{L}}_{\tau }\left(
\begin{array}{lllllllll}
t &  & s_{1} &  & s_{2} &  & \tau &  & t \\
& r_{1} &  & l_{1} &  & r^{\prime } &  & l^{\prime } &
\end{array}
\right)  \nonumber \\
&&+\frac{2}{2!}\mathcal{\hat{L}}_{\tau }\left(
\begin{array}{lllll}
\tau &  & s_{2} &  & \tau \\
& l^{\prime } &  & r^{\prime } &
\end{array}
\right) \cdot \mathcal{\hat{L}}_{\tau }\left(
\begin{array}{lllll}
t &  & s_{1} &  & t \\
& r_{1} &  & l_{1} &
\end{array}
\right)  \nonumber \\
&&-\frac{2}{2!}\mathcal{\hat{L}}_{\tau }\left(
\begin{array}{lllllllll}
t &  & \tau &  & s_{2} &  & s_{1} &  & t \\
& r^{\prime } &  & l^{\prime } &  & r_{1} &  & l_{1} &
\end{array}
\right) .  \nonumber \\
&&  \label{eq:loopk2}
\end{eqnarray}
Note that all contour-ordering $\{T_{\tilde{p}_{i}}\}$ of the leading and
the second order perturbations have been taken into account in Eqs. (\ref
{eq:loopk0},\ref{eq:loopk2}) combining with their duality, respectively.

In addition, we can express each order of the kernel in terms of the loop
operators associated with the irreducible diagrams, for instance,
\begin{eqnarray*}
&&K^{(0)}(t-\tau )*\rho _{S}(\tau ) \\
&=&Q_{S}(t-t_{0})\mathcal{\hat{L}}_{ir,\tau }\left( t,\tau ,t\right)
\mathcal{J}(\hat{q},\hat{q}^{+};\vec{\alpha},\vec{\gamma})\Big|_{\vec{\gamma}%
=\vec{\gamma}^{\prime }=\{\vec{\alpha}\}=0},
\end{eqnarray*}
\begin{eqnarray}
&&K^{(2)}(t-\tau )*\rho _{S}(\tau )  \nonumber \\
&=&Q_{S}(t-t_{0})\Big\{\mathcal{\hat{L}}_{in,\tau }\left[
s_{1},s_{2},s_{1}\right] \cdot \mathcal{\hat{L}}_{ir,\tau }\left( t,\tau
,t\right)  \nonumber \\
&&+\mathcal{\hat{L}}_{ir,\tau }\left( \tau ,s_{2},\tau \right) \cdot
\mathcal{\hat{L}}_{ir,\tau }\left( t,s_{1},t\right) -\mathcal{\hat{L}}%
_{ir,\tau }\left( t,s_{1},s_{2},\tau ,t\right)  \nonumber \\
&&-\mathcal{\hat{L}}_{ir,\tau }\left( t,s_{1},\tau ,s_{2},t\right) -\mathcal{%
\hat{L}}_{ir,\tau }\left( t,\tau ,s_{1},s_{2},t\right) \Big\}  \nonumber \\
&&\times \mathcal{J}(\hat{q},\hat{q}^{+};\vec{\alpha},\vec{\gamma})\Big|_{%
\vec{\gamma}=\vec{\gamma}^{\prime }=\{\vec{\alpha}\}=0}.
\end{eqnarray}
Therefore, we can resum all perturbation orders of $K(t-\tau )*\rho
_{S}(\tau )$ according to the irreducible diagrams.

\end{document}